\documentclass[usenatbib]{mn2e}
\def\lapp{\ifmmode\stackrel{<}{_{\sim}}\else$\stackrel{<}{_{\sim}}$\fi}
\def\gapp{\ifmmode\stackrel{>}{_{\sim}}\else$\stackrel{<}{_{\sim}}$\fi}
\title[ Braking Indices of Magnetars ]
{Constraining the Braking Indices of Magnetars}

\author[Z. F. Gao et al.]
      {Z. F. Gao$^{1, 2}$,\thanks{E-mail: zhifugao@xao.ac.cn}
   X.-D. Li$^{3}$, N. Wang$^{1}$ \thanks{E-mail: na.wang@xao.ac.cn}, J. P. Yuan$^{1}$, P. Wang$^{4}$, Q.H. Peng$^{3}$, Y.J. Du$^{5}$, \\
$^{1}$ Xinjiang Astronomical Observatory, Chinese Academy of Sciences, 150, Science
1-Street, Urumqi, Xinjiang, 830011, China\\
$^{2}$ Key Laboratory of Radio Astronomy, Chinese Academy of Sciences, West Beijing Road, Nanjing, Jiangsu 210008, China\\
$^{3}$ Department of Astronomy and Key Laboratory of Modern
Astronomy and Astrophysics, Nanjing University, Jiangsu 210046, China\\
$^{4}$ National Astronomical Observatories, Chinese Academy of Sciences, Beijing 100012, China\\
$^{5}$ Qian Xuesen Laboratory of Space Technology, Beijing 100094, China}
\date{\today}

\pagerange{\pageref{firstpage}--\pageref{lastpage}} \pubyear{2014}
\usepackage{epsfig}   
\usepackage{natbib}  
\usepackage{lscape}   
\usepackage{booktabs}   
\usepackage{threeparttable} 
\usepackage{amsmath}

\begin{document}

\label{firstpage}

\maketitle

\begin{abstract}
Due to the lack of long term pulsed emission in quiescence and the
strong timing noise, it is impossible to directly measure the
braking index $n$ of a magnetar. Based on the estimated ages of
their potentially associated supernova remnants (SNRs), we estimate
the values of the mean braking indices of eight magnetars
with SNRs, and find that they cluster in a range of $1\sim$42\textbf{}.
Five magnetars have smaller mean braking indices
of $1<n<3$, and we interpret them within a combination of
magneto-dipole radiation and wind aided braking, while the larger mean
braking indices of $n>3$ for other three magnetars are
attributed to the decay of external braking torque, which might be
caused by magnetic field decay. We estimate the possible wind
luminosities for the magnetars with $1<n<3$, and the dipolar
magnetic field decay rates for the magnetars with $n>3$ within the
updated magneto-thermal evolution models.  Although the constrained range of the
magnetars' braking indices is tentative, due to the
uncertainties in the SNR ages, which come from distance
uncertainties and the unknown conditions of the expanding shells,
 our method provides an effective way
to constrain the magnetars' braking indices if the
measurements of the SNRs' ages are reliable, which can be improved by future observations.
\end{abstract}

\section{Introduction} \label{sec:intro}

Pulsars are commonly recognized as magnetized neutron stars (NSs), sometimes have
been argued to be quark stars \citep{zyl06, xu07, lgx13}.  A pulsar's secular
spin-down is mainly caused by its rotational energy losses due to
electromagnetic radiation, particle winds, star-disk interaction, intense
neutrino emission or gravitational radiation\,(e.g., Peng et al. 1982;
Harding et al. 1999; Alpar \& Baykal 2006; Chen \& Li 2006).
An important and measurable quantity closely related to a pulsar's  rotational evolution
is the braking index $n$, defined by assuming that
the star spins down in the light of a power law
\begin{equation}
  \dot{\Omega}=~-K \Omega^{n}~,
  \label{1-equation}
\end{equation}
where $\Omega$ is the angular velocity, $\dot{\Omega}$ is the derivative of $\Omega$, and $K$ is a
proportionality constant. When the magneto-dipole radiation (MDR)
solely causes the pulsar spin-down, the braking index is predicted to be $n=3$. Note that
this constant value of $n=3$ holds only if the torque is proportional to $\Omega^{3}$,
which is the case for the magnetic torque with $K$ keeping constant over time\,
(Blandford \& Romani 1988). However, in general $n$ can vary over time because variations
in any of these quantities may result in
departures from this predicted value of 3.

The standard way to define the braking index is
\begin{equation}
  n\equiv \frac{\Omega\ddot{\Omega}}{\dot{\Omega}^{2}}=
  ~\frac{\nu\ddot{\nu}}{\dot{\nu}^{2}}= 2-\frac{P\ddot{P}}{\dot{P}^{2}},
  \label{2-equation}
\end{equation}
where $\ddot{\Omega}$ the second derivative of $\Omega$,
$\nu=\Omega/2\pi$ is the spin frequency, and $P=1/\nu$ the spin
period (see Manchester \& Taylor 1977, and references therein). Be
cautious that the measured value of $n$ strongly depends on a
certain time-span, and any variation in the braking torque
will give a time-varying $n(t)$. The braking index also reveals
some simple physical expectation. For example, $n=5$ may point to
gravitational-wave-emission\footnote{For a normal radio pulsar,
gravitational wave radiation can be only efficient during the first
minutes/hours of the life (Manchester \& Taylor 1977).} as the
dominant spin-down mechanism, while in the case of $n=1$,
particle-wind radiation may be the cause of the pulsar spin down.

For the majority of pulsars, it is extraordinarily
difficult to measure their stable and sensible braking
indices $n$, from pulsar timing-fitting by using Eq.(2).  The main
reason is that the second derivative of $\nu$
is very small and its detection is affected
by the rotational irregularities, such as glitches, an abrupt
increases in $\nu$ and $|\dot{\nu}|$ followed by relaxations\,(Yuan
et al. 2010), and timing noise (Zhang \& Xie 2013; Kutukcu
\& Ankay 2014; Haskell \& Melatos 2015).

By analyzing the timing residuals of 366 non-recycled
pulsars over the past 36 yrs, Hobbs et al. (2010) measured their
braking indices and showed that they rang from $-287986$ to $+3624$.
Unfortunately, the observed $\ddot{\nu}$ for the majority of these
pulsars with $t> 10^{4}$\, yrs are dominated by the amount of timing
noise present in the residuals and the data span. As noted in Pons
et al. (2012), any small relative variation of the torque will
produce large variation in $n$ for old pulsars.
In general, timing noise is ubiquitous among pulsars. Only for young radio pulsars, which spin
down quickly and have stable first and second spin-period derivatives, their stable and reliable braking
indices have been measured (shown as in Table 1), though these sources experience few small or large glitches, the post-
glitch recoveries dominate their timing behavior (Hobbs et al. 2010).
Ten sources have
positive braking indices below the canonical value of 3.
\begin{table}
\caption{ Braking indices of 11 young radio pulsars. }
\begin{tabular}{ccc}
\\
\hline 
Source Name & n & Reference\\
 \hline
B0531$+$21(Crab)    &2.51(1)   & Lyne et al. 1993 \\
J0537$-$6910   &$-$1.5(1)      & Middleditch et al. 2006  \\
B0540$-$69     &2.140(9)  & Ferdman et al. 2015 \\
B0833$-$45(Vela)     &1.4(2)      & Lyne et al. 1996 \\
J1119$-$6127   &2.91(5)   & Weltevrede et al. 2011 \\
B1509$-$58     &2.839(1)  &  Livingstone et al. 2007 \\
J1734$-$3333   &0.9(2)    & Espinoza et al. 2011 \\
 B1800$-$21$^{*}$     & 2(1)       & Espinoza 2013\\
B1823$-$13$^{*}$    &2(1)        & Espinoza 2013 \\
J1833$-$1034   &1.857(6)  & Roy et al. 2012 \\
J1846$-$0258   &2.65(1)   & Livingstone et al. 2007 \\
\hline\\
\end{tabular}\\
\label{tb:braking index 1}
\noindent{$*$: For pulsars B1800$-$21 and B1823$-$13,
the values of $n=2$ quoted in Espinoza (2013) are preliminary values of braking indices, with
error bars of the order of 1 (Espinoza, private communication).}
\end{table}

The very small braking index of $n=0.9\pm0.2$ (for PSR J1734$-$3333) and
$n=-1.5(1)$ (for PSR J0537$-$6910)
could be consistent with the reemergence of magnetic fields after a deep
submergence into the star's crust (Muslimov \& Page 1995; Ho 2011; Vigan\`{o} \& Pons 2012; Gourgouliatos \& Cumming 2014).
According to the magneto-thermal evolution model proposed by Pons et al. (2012),
a reemergence or a temporal rise of the crustal field, due to Ohmic dissipation and
Hall drift, could in principle provide any braking-index value smaller than 3.
Braking indices reaching values of $\pm10^{4\sim5}$ were also
reported\,\citep{mht05}.  These observations could be contaminated by
unsettled glitches or timing noise\,\citep{ab06}.  Other
interpretations for such large $|n|$ have also been proposed
 \citep{bt10, bbk12, pvg12, zx12}.

Anomalous X-ray pulsars (AXPs) and soft gamma repeaters (SGRs) are
usually regarded as magnetars (Duncan \& Thompson 1992;
Thompson \& Duncan 1996). They are young NSs endowed with large
surface dipolar magnetic fields. The soft X-ray emission in quiescence, bursts and flares of magnetars
is supposed to be fueled by the rearrangement and decay of the
internal magnetic fields (e.g., Gao et al. 2011,
2012, 2013)\footnote{The internal magnetic field of a magnetar
could be a mixture of toroidal and poloidal magentic fields, and the
toroidal component dominates (Mereghetti 2008; Olausen \& Kaspi 2014).}.

It is commonly believed that pulsars, including traditional radio pulsars and
magnetars, are produced in core-collapse supernova explosions. Supernova
remnants (SNRs) are the expanding diffuse gaseous nebulas resulting from
the explosions of massive stars. SNRs can provide direct insights into
the NS progenitor models and the explosion mechanisms, and thus have been studied
by many groups (e.g., Vink \& Kuiper 2006; Tian et al. 2007; Martin et al. 2014; Zhu et al. 2014).

To date, we have obtained available information of 28 known magnetars: 14
SGRs (11 confirmed), and 14 AXPs (12 confirmed). Among
these sources, only 4  magnetars (1E 1547.0$-$5408, PSR J1622$-$4950, SGR
J1745$-$2900 \citep{ysw15} and XTE 1810$-$197) were discovered to emit radio waves.  With the exception of SGR 0418$+$5729,
all known Galactic magnetars lie within 2$\degr$ of the Galactic Plane, consistent
with a population of young objects, see ``The McGill
Magnetar Catalog'' for relevant information. An online version of the catalog is located
at the web site http://www.physics.mcgill.ca/~pulsar/magnetar/main.html,
\citep{ok14}.

The study of the braking indices of magnetars can provide a wealth
of information about the magneto$-$thermal evolution and the
spin-down evolution of these objects. However, due to the
lack of long term pulsed emission in quiescence and the strong
timing noise, we have not yet measured the values of $n$ for
magnetars.  Observations hint that some magnetars have associations
with clusters of massive stars. Recently, Tendulkar et al.
(2012) measured the linear transverse velocities, then estimated the
kinetic ages of SGR 1806$-$20 and SGR 1900$+$14 assuming that the
magnetars were born in star clusters. Using the following implicit
equation
\begin{equation}
n=1+\frac{P}{T\dot{P}}\left[1-\left(\frac{P_{i}}{P}\right)^{n-1} \right],
\label{3}~
\end{equation}
and assuming $P_{i}/P\ll 1$, the authors estimated $n$ to be
1.76$^{+0.65}_{-0.24}$ for SGR 1806$-$20 and 1.16$^{+0.04}_{-0.07}$
for SGR 1900$+$14. Here $T$ denotes the kinematic age of the magnetar
(the time taken to move from the cluster to present position) and $P_{i}$ is
the initial spin period at $t=0$ (Tendulkar et al. 2012).
Although the braking indices of magnetars have been
investigated by some authors (e.g., Magalhaes et al. 2012), the work of specifically constraining the
braking \textbf{indices of  magnetars} has not appeared so far. In this paper,
we try to constrain the braking indices of magnetars from
the ages of their SNRs.  In Sec. 2,
we describe the potential magnetar/SNR associations. In Sec. 3, we
present constrained values of $n$ for eight magnetars with associated SNRs, and
possible explanations. In Sec. 4, we give a summary and discussion.

\section{Potential magnetar/SNR Associations  \label{sec:comments}}
It was predicted that about $10\%$ of Galactic supernova explosions
may lead to magnetars \citep{kfm94}.  Observations hint that
about one third of the magnetars are likely to be associated with SNRs,
suggestive of an origin of massive star explosions. Currently, the
remnants ages are estimated based on the measurements of the shock
velocities, and/or the X-ray temperatures and/or other quantities.
The SNR evolution can be schematically divided into three successive phases:

(i) An ejecta-dominated phase.  A shock wave created by the supernova explosion
has a free expansion velocity of $5000-10000$~km~s$^{-1}$(Woltjer 1972).

(ii) A Sedov-Taylor phase. The radiative losses from a SNR in this adiabatic
expansion stage are dynamically insignificant and can be neglected.

(iii) A radiation-dominated snowplow phase. When the temperature
drops below about $10^{6}$\,K, radiation becomes important, and energy is lost
mainly through radiation. This phase ends with the dispersion of the envelope as
its velocity falls to about 9\,km~s$^{-1}$ (Bandiera 1984).

Sedov (1959) showed that there is a self-similar solution for
the adiabatic expansion phase (i.e., the structure of a SNR keeps constant within
this stage), and firstly applied this solution
to estimate the age of SNRs in this phase, i.e., the Sedov age. The Sedov age estimates
rely on many assumptions, but ultimately on the SNR size and the SNR temperature.
If a SNR is too young (when the expansion is free), or too old (when radiation
energy losses brake the expansion), the Sedov expansion is no longer applicable.

Many authors (e.g., Ostriker \& McKee 1988; Truelove \& McKee 1999;
Gaensler \& Slane 2006) subsequently investigated the Sedov ages of SNRs.  Based
on $XMM-Newton$ observations, Vink \& Kuiper (2006) employed the Sedov evolution,
and estimated the supernova explosion energies $E$, and the SNR ages $t_{\rm SNR}$ by
\begin{equation}
 2.026Et_{\rm SNR}^{2}= R_{s}^{5}\rho_{0},
  \label{4-equation}
\end{equation}
where $\rho_{0}$ is the density of the interstellar medium (ISM), and $R_{s}$ is the
shock radius, which approximates to the SNR radius $R_{\rm SNR}$.
The shock velocity $v_{s}$ can be obtained by
\begin{equation}
 kT=\frac{3}{16}\times 0.63m_{p}v_{s}^{2},
  \label{5-equation}
\end{equation}
where $T$ is the post-shock plasma temperature, $m_{p}$ is the
proton mass, and the number 0.63 gives the mean particle
mass in units of $m_{p}$ for a fully ionized plasma.
From Eq.(5), we can obtain the value of $v_{s}$, given the
measured temperature $T$. The age of the remnant can be estimated from
the shock velocity similarity solution
\begin{equation}
 v_{s}=\dot{R}_{\rm SNR}=\frac{2R_{\rm SNR}}{5t_{\rm SNR}}.
  \label{6-equation}
\end{equation}
Inserting Eq.(5) into Eq.(6), we get a convenient expression
\begin{equation}
t_{SNR}\approx 435\times R_{\rm SNR}T^{-1/2}~~{\rm yrs}~,
  \label{7-equation}
\end{equation}
where $R_{\rm SNR}$ and $T$ are in units of pc and keV, respectively
\footnote{The propagated error of $R_{\rm SNR}$ is given by $\Delta
t_{\rm SNR}=435\times(\frac{\Delta^{2}R_{\rm SNR}}{T}+\frac{R^{2}_{\rm SNR}
\Delta^{2}T}{4T^{3}})^{1/2}$\,yrs.}.
With respect to the potential magnetar/SNR associations, Gaensler (2004) proposed
important criteria, which include evidence for interaction, consistent
distances/ages, inferred transverse velocity, proper motion and
the probability of random alignment. However, due to the observational limit,
these criteria cannot always be applicable. The judgement as to whether a
potential magnetar/SNR association meets these requirements ultimately depends on
observational evidences. As pointed out by Gaensler (2004), one basic indication
of the likelihood of an association is the chance coincidence probability
between a magnetar and its host SNR.

In the following, we briefly review the measured and derived
parameters for the \textbf{eleven} magnetars and their associated SNRs.  All these SNRs
have kinetic distances derived from their associations with other objects
(e.g., molecular clouds) having known distances.
\begin{enumerate}
\item 1E 1841$-$045 (hereafer 1E 1841)~~The error box of this AXP lies within the
SNR Kes 73\,(G27.4$+$0.0), which is nearly circular in X-rays with an angular diameter of $\theta=4.5^{'}$ (Helfand et al. 1994; Gotthelf \& Vasisht 1997).
The probability of chance alignment was estimated to be $1\times10^{-4}$ for this association
(Vasisht \& Gotthelf 1997; Gaensler 2004). Using the galactic
circular rotation model, Tian \& Leahy (2008) gave a range of revised distance
to Kes 73 about 7.5$-$ 9.8\,kpc, which implies a shock
radius $R_{\rm SNR}\approx4.9-6.4$\,pc. Then the Sedov age of Kes 73 was
estimated to be $500-1000$\,yrs for standard explosion energy ($E=10^{51}$\,erg)
and ISM $n$=1\,cm$^{-3}$ (Tian \& Leahy 2008).

\item SGR 0526$-$66 (hereafer SGR 0526)~~This SGR could be physically associated with the remnant N49
(Evans et al. 1980) located in the Large Magellanic
Cloud (LMC) at a distance of $50$ kpc and interacting with a
molecular cloud. The deep \textit{Chandra} observation indicated the gas
temperature in the shell region to be $T$=0.57\,keV for N49 (Park et al. 2012). Assume an angular distance
of 35$^{''}$ between the Shell region and the geometric center of N49, the SNR radius is
estimated to be 8.5 pc, then the Sedov age of the SNR is 4800 yrs (Park et al. 2012).
However, the association between this SGR and N49 is controversial (e.g., Gaensler et al. 2001; Klose et al. 2004; Badenes et al. 2009), and more evidence is needed to testify this potential association.

\item SGR 1627$-$41 (hereafer~SGR 1627)~~This SGR is associated with the G337.0$-$0.1
(Hurley et al. 1999), and is located on the edge of massive molecular clouds.
The angular size (radius of $\sim45^{''}$) of G337.0$-$0.1 implies a radius
of 2.4 pc at a distance of 11.0\,kpc (Sarma et al. 1997).  The spurious
probability of the SGR/SNR association was estimated to $\sim 5\%$ (Smith et al.
1999). The age of the SNR was estimated to be 5000 yrs based on the Swedish-ESO
Submillimeter Telescope observations (Corbel et al. 1999).

\item SGR 0501$+$4516 (hereafer SGR 0501)~~This SGR could be associated with SNR
HB9 (also known as G160.9$+$2.6)(Gaensler \& Chatterjee 2008) with
an angular diameter of 128$^{'}$ (Leahy \& Aschenbach 1995), but
this association is not confirmed by observations (Barthelmy et al. 2008).
The centrally brighten X-ray emission with $T\sim0.82$\,keV from HB9
can be explained by self-similar evolution in a cloudy
interstellar medium proposed by White \& Long (1991)
(WL91). Applying the WL91 evaporative cloud model, Leahy \& Aschenbach (1995)
(LA95) gave a distance of $d=$1.5\,kpc (corresponding to $R_{\rm SNR}\sim28$\,pc) and an age
of 7700\,yrs for the SNR. Based on the H I absorption measurement, Leahy \& Tian (2007)\,(LA07) obtained a distance of $d= 0.8\pm$0.4\,kpc (the mean radius is 15\,pc at $d=0.8$\,kpc), and an age of 4000$-$7000\,yrs for the SNR within WL91 evaporative cloud model. In the following, we simultaneously consider both the results of LA95 and LT07, i.e., the distance is assumed to be  $0.4-1.5$\,kpc, and the age for the SNR varies between 4000 and 7700 yrs.

\item PSR J1622$-$4950 (hereafer PSR J1622)~~The positional coincidence of this AXP
with the center of G333.9$+$0.1 with an angular diameter about 12$^{'}$
suggests a possible magnetar/SNR association (Anderson et al. 2012).
G333.9$+$0.1 has a radius of 18 pc at the distance of 9 kpc, and is in
a low density medium, suggesting the Sedov-Taylor expansion. (Anderson et
al. 2012). Since no X-ray emission was detected from the SNR in the $XMM$-$Neuton$
observation (Anderson et al. 2012), the upper limit on the
count-rate was estimated as 0.04 counts s$^{-1}$ utilizing standard errors
and roughly accounting (Romer et al. 2001). Assuming
a thin-shell morphology and the Sedov-Taylor solution (Sedov 1946,
1959; Taylor 1950), the upper limit on the count-rate implies an upper
limit on the preshock ambient density to be $n_{0}$=0.05\,cm$^{-3}$  for an explosion energy of $10^{51}$\,erg. Then the upper-limit for the SNR age was estimated to be 6\,kyrs (Anderson et al. 2012).


\item 1E 2259$+$586 (hereafer 1E 2259)~~This AXP lies within 4$^{'}$ of the geometric
center of CTB 109 (Fahlman \& Gregory 1981) with an angular radius of
$18.^{'}5\pm 1.^{'}0$ as estimated from the $XMM$-$Neuton$ EPIC images by
Sasaki et al (2004). The probability of chance alignment is about $10^{-4}$
for this association (Gaensler 2004). There are some more recent
estimates of the distance to CTB 109 and the AXP.  Kothes
\& Foster (2012) collected all the observational limits of the
distance (Kothes et al. 2002; Durant \& van Kerkwijk 2006; Tian et al.
2010), and estimated a consensus distance of $3.2\pm0.2$\,kpc to CTB 109,
which places the SNR inside the Perseus spiral arm of the Milky Way.
Castro et al. (2012) reported the detection of a GeV source
with $Fermi$ at the position of CTB 109. By simulating in the
CR(cosmic rays)-Hydro-NEI (nonequilibrium ionization) model of Ellison et al.(2007),
Castro et al. (2012) derived an age of the SNR to be $8.5-15$\,kyrs, which is
superior to the previous age estimates. Based on the $Chandra$ data and the Sedov-Taylor
solution, Sasaki et al. (2013) estimated an age of $10-20$\,kyr for the SNR
from a two-component model for the X-ray spectrum. Sasaki et al. (2013)
also concluded that their (new) $Chandra$ study is consistent with the $Fermi$ results by Castro et al. (2012).

\item CXOU\,J171405.07$-$381031\,(hereafer CXOU J1714)~ Halpern \& Gotthelf
(2010a) firstly suggested an association of this AXP with CTB 37B having a size of $17^{'}$
(see the SNR Catalog\footnote{at http://www.physics.umanitoba.ca/snr/SNRcat/}),
and a distance of $10.2\pm3.5$\,kpc, Caswell et al. 1975). This association is possible because the first TeV source HESS J1713$-$381
is hosted in CTB 37B, and is coincident with the AXP CXOU J1714 (e.g., Sato et al.\,2010; Halpern
\& Gotthelf 2010b).  Horvath \& Allen (2011) gave an extensive review of this association
and the SNR age. Clark \& Stephenson (1975) obtained age estimates for both
CTB 37A and CTB 37B of $\sim1500$\,yrs, and claimed that either source could be
SN 393. Such an association is possible, but the SNRs may be considerably older
(Downes 1984).  Assuming an electron-ion equilibrium, Aharonian et al. (2008) gave
a Sedov age of $\sim 4900$\,yrs for CTB 37B. The inferred age for the SNR may be
decreased to $\sim2700$ yrs due to a higher ion temperature and efficient cosmic
ray acceleration (Aharonian et al. 2008). Note that the electron-ion equilibrium
is usually not achieved in young SNRs. By introducing ``magnetar-driven'' injected
energy, Horvath \& Allen 2011 obtained an age of $\sim 3200$ yrs (see Fig.1
in their work) for CTB 37B, but the explosion energy and the ISM density were always
fixed to be $10^{51}$\,erg and 1\,cm$^{-3}$, respectively. Based on the $Suzaku$ observations,
Nakamua et al.\,(2009) obtained an ionization age of the thermal plasma
associated with CTB 37B as 650$^{+2500}_{-300}$\,yrs, which is consistent with the
tentative identification of the SNR with SN 393.

\item Swift\,J1834.9$-$0846 (hereafer Swift J1834)~~This transient magnetar is
located at the center of the radio SNR W41 and the TeV source HSS
J1844$-$087 (Kargaltsev et al. 2012).  Such a cental location of
Swift J1834.9$-$0846 certainly supports the association between
this SGR with W41. H I observations from the VLA Galactic Plane
Survey gave a close distance of $4\pm0.2$\,kpc, and
an average radius $\sim19\pm1$\,pc for W41 (Tian et al. 2007).
The authors estimated an age of about $60$\,kyr, using a Sedov model with
$E\sim0.75\times10^{51}$\,erg, and an age of about $200$\,kyrs from
complete cooling expansion (Cox 1972).

\begin{table*}
\scriptsize
\begin{center}
\centering
\begin{minipage}{178mm}
\caption{Persistent parameters for eight magnetars and their SNRs. Column one to ten are source name, spin period, spin-period derivative, surface dipole magnetic field, characteristic age, possibly associated SNR, SNR distance, SNR radius, SNR age and reference, respectively. Data in Columns one to six are cited from Table 1 of McGill SGR/AXP Online Catalogue, up to Feb. 6, 2015. Data in Columns seven to ten are mainly cited from the SNR Catalog at http://www.physics.umanitoba.ca/snr/SNRcat/ and from the Coolers Catalog at http://www.neutonstarcooling.info/references.html. In case where the timing properties vary with either phase or time, or there exists multiple recently measured values, the entries are marked with an asterisk ($*$) and compiled separately in Table 3.}
\label{tb:parameters}
\begin{tabular}{cccccccccc}

\scriptsize
\\
\hline 
Name &$P$ & $\dot{P}$ &$B_{\rm d}$ &${\tau}_{\rm c}$ &SNR  & $d_{\rm SNR}$& $R_{\rm SNR}$  & $t_{\rm SNR}$ & Ref. \\
 &(s) & $10^{-11}\,\rm s/s$  & $10^{14}\,\rm Gauss$ & kyrs& &kpc & pc &  kyrs  & \\
 \hline
1E 1841 &11.788978(1) & 4.092(15)&7.0 &4.6  &Kes73 & 7.5$-$9.8 &4.9$-$6.4  &0.75$\pm$0.25  &[1-3]   \\
SGR 0526$^*$  & 8.0544(2)&3.8(1) &5.60(7) &3.36(9) &N49 & 50 & 8.5   & 4.8    &[4-5]  \\
SGR 1627 & 2.594578(6) &1.9(4) &2.25(9) &2.2(5) &G337.0$-$0.1  &11.0 & 2.4   &5.0  &[6-8]  \\
SGR 0501 & 5.76209695(1)&0.594(2) &1.9 &15 & HB9  & 0.4$-$1.5  & 7.5$-$28 &5.85$-$1.85 &[9-11] \\
PSR J1622$^*$   &4.3261(1) &1.7(1) &2.74(8)&4.0(2) &G333.9$+$0.0 & 9& 18 &$<6.0$  &[12] \\
1E 2259  &6.9790427250(15) &0.048369(6) &0.59 &230 &CTB 109  &3.0$-$3.4$^{\dag}$ &15$-$17$^{\dag}$ &15$\pm$5     &[13-15]\\
CXOU J1714$^*$  &3.825352(4) &6.40(5) &5.0  &0.95(1) &CTB 37B &10.2 &25.2  &1.75$\pm$1.4 &[16-18] \\
Swift J1834 &2.4823018(1) &0.796(12) &1.42(1)&4.9 &W41  &3.8$-$4.2 &18$-$20 &130$\pm$70      &[19-20] \\
\hline\\
\end{tabular}\\
\noindent{Footnotes:$^\dag$ In Sasaki et al. 2013, a very close value of $d=2.9-3.5$\,kpc was adopted, because the distance to the density peak of H I associated with arm was permitted to vary between 2.9\,kpc and 3.6\,kpc, although the distance to the SNR is mainly similar at 3.2 kpc, according to Kothes \& Foster 2012. \\
Refs: [1] Helfand et al. 1994 [2] Gotthelf \& Vasisht 1997 [3] Tian \& Leahy 2008 [4] Evans et al. 1980 [5] Park et al. 2012  [6] Sarma et al. 1997 [7] Hurley et al. 1999 [8] Corbel et al. 1999 [[9] Gaensler \& Chatterjee 2008 [10] Leahy \& Aschenbach 1995 [11] Leahy \& Tian 2007 [12] Anderson et al. 2012 [13] Kothes et al. 2002 [14] Kothes \& Foster 2012 [15] Sasaki et al. 2013 [16] Caswell et al. 1975 [17] Aharonian et al. 2008 [18] Nakamua et al. 2009  [19] Tian et al. 2007 [20] Kargaltsev et al. 2012 }
\end{minipage}
\end{center}
\end{table*}

\item  The AXP 1E 1547$-$5408 lies within
$30^{''}$ of the center of G327.24$-$0.13 with a size of $5^{'}$ in
radio (see in the SNR Catalog). The physical association of theses
two sources was firstly proposed by Gelfand \& Gaensler (2007), and
was established by other observations (e.g., Joseph et al. 2007;
Vink \& Bamba 2009), and the distance to the SNR is quite uncertain.
Unfortunately, since there is no X-ray emission from G327.24$-$0.13,
we cannot estimate the age of this SNR by its Sedov evolution, and
the published age estimate for this SNR from other methods has not
appeared so far. Thus, we cannot estimate the braking index of 1E
1547$-$5408 Another magnetar AX J1845.0$-$0258 lies with the
5$^{'}$ diameter SNR G29.6$+$0.1, and the probability of chance
alignment between the source and the SNR is shown to be $1.6\times
10^{-3}$ (Gaensler et al. 1999). However, due to a long-term burst
behaviors, there is a lack of detection of the period
derivative or dipole magnetic field for AX J1845.0$-$0258, so the
braking index of AX J1845.0$-$0258 also cannot be estimated.  The
newly identified magnetar SGR 1935$+$2154 was proposed to be
coincident with the Galactic SNR G57.2$+$0.8 (Gaensler et al. 2014;
Sun et al. 2011). To date, no reliable age or distance estimate is
available for G57.2$+$0.8, and for the evaluation of the braking
index of SGR 1935$+$2154. Thus, the three magnetars
above will not be considered in the following.
\end{enumerate}

\begin{table}
\centering
\caption{Variable or alternative values of persistent parameters for three magnetars with SNRs.
All data are cited from Table 2 of McGill SGR/AXP Online Catalogue at
http://www.physics.mcgill.ca/$\sim$pulsar/magnetar/table2.html, up to Jan 10, 2015.}
\label{tb:parameters2}
\begin{tabular}{ccccc}
\\
\hline 
Name &$P$ & $\dot{P}$ &$B_{\rm d}$  &${\tau}_{\rm c}$   \\
 &(s) & $10^{-11}\,\rm s/s$  & $10^{14}\,\rm Gauss$ &  kyrs  \\
\hline\\
SGR 0526  & 8.0470(2)&6.5(5) &7.3(3) & 2.0(1)  \\
PSR J1622  & 4.3261(1)&0.94$-$1.94 &2.5(4) &5(2).5  \\
CXOU J1714 &3.823056(18) &5.88(8) &4.80(4) &1.03(1) \\
             $----$ &3.824936(17) &10.5(5) & 6.4(1)&0.58(3)\\
\hline\\
\end{tabular}\\
\end{table}

Now, we list the parameters for the eight magnetars and their SNRs.
 All the SNRs in Table 2 have published
age estimates. Among these objects, a few SNRs have several possible values
of $t_{\rm SNR}$, measured with different methods at different times,
 we select the age $t_{\rm SNR}$ that is the latest and most
reliable (e.g., $t_{\rm SNR}\sim (15\pm 5)$\,kyr for CTB 109, as shown in Table 2).
For convenience, we denote each SNR age ($t_{\rm SNR}$) in the same
form of ``main value $+$ age error''.
With the exception of SNR W41 ($t_{\rm SNR}\sim 130\pm70$\,kyrs), all the
ages of the magnetar SNRs are mainly distributed within a rather narrow range of
$1-20$\,kyrs.

\section{Constraining magnetar braking indices   \label{sec:predicte}}
\subsection{Constrained values of $n$ for magnetars  \label{subsec:index}}
Assuming a constat value of $n=3$, the characteristic age of a NS is
defined as $\tau_{\rm c}=P/2\dot{P}$.  As we know, $\tau_{\rm c}$ is
usually to be used to as an approximation to its real age, with which
it coincides only if the current spin period was much larger than the
initial value, the magnetic-field strength $B$, the moment of inertia, and the angle between the magnetic and rotational axes have been constant
during the entire NS life. By integrating Eq.(1), we get the inferred age as
\begin{eqnarray}
&&t=\frac{P}{(n-1)\dot{P}}~\left[1- \left(\frac{P_{i}}{P}\right)^{n- 1}\right]~~~~\left(n\neq 1\right),\nonumber\\
&&t=2\tau_{\rm c}~\rm{\ln} \left(\frac{P}{P_{i}}\right)~~~~~~~~~~~~~~~~~\left(n= 1\right),
\label{8-equation}
\end{eqnarray}
given a constant braking index $n$, where $t$ is the
inferred age (e.g., Manchester et al. 1985; Livingstone et al. 2011).

In order to constrain the values of $n$ for magnetars,
we make the following three assumptions:
\begin{enumerate}
\item Firstly, we assume that a NS's braking index keeps constant since its birth.
This is a strong necessary assumption when estimating the age of a
very young NS. In deed, one cannot derive an analytical expression, as Eq. (8), to estimate the age of a NS, if we consider
either $K$ or $n$ in Eq. (1) is a time-dependent quantity.  The
measured braking index is obviously influenced by the variation in
the braking torque during the observational intervals. Nevertheless,
the derived braking index can provide useful constraints on the
realistic one.

\item Secondly, we assume that the age of a SNR is the age of a NS. Zhang \& Xie (2012)
suggested that the physically meaningful criterion to estimate the
true age of a  NS is the NS/SNR association. This suggestion is
surely in the same way applicable to magnetars with hosted SNRs.

\item Finally, we assume that the initial spin-period of a magnetar is far less than the
current spin-period, i.e., $P_{i}\ll P$, because magnetars are supposed to be formed from
rapidly rotating NSs (e.g. Duncan \& Thompson 1992; Thompson \& Duncan 1996).
\end{enumerate}
Base on the above assumptions, the inferred age in Eq. (8) can be estimated as
\begin{equation}
t\approx~\frac{P}{(n- 1)\dot{P}}= -\frac{\nu}{(n-1)\dot{\nu}}~,
\label{9-equation}
\end{equation}
for $\nu_{i} \gg \nu$.
Inserting the relation $t= t_{\rm SNR}$ into Eq. (9) gives
\begin{equation}
n\approx 1-\frac{\nu}{\dot{\nu}t_{\rm SNR}}=1+\frac{P}{\dot{P}t_{\rm SNR}}~.
\label{10-equation}
\end{equation}
It should be noted that $t$ in Eq.(9) is the upper limit of the
inferred age due to an approximation in Eq.(8), and $n$ denotes the
mean braking index of a magnetar since its formation due to the
assumption of a constant braking index. In the following Sections,
for the sake of convenience, the mean braking index will be called
``the braking index" in short.

We can calculate the values of $n$ for eight magnetars
with SNRs by combining the data in Tables 2 and 3 with Eq.(10),
which are presented in Table 4.
\begin{table}
\caption{Constrained values of $n$ for the eight megnetars with SNRs. The alternative
braking indices are marked with an asterisk($*$), and calculated from the data in
Table 3. }
\begin{tabular}{ccc}
\\
\hline 
Source  &$n$ & Timing Reference.  \\
 \hline
1E 1841      &13$\pm$4  & Dib \& Kaspi 2014       \\
SGR 0526     & 2.40$\pm$0.04 & Tiengo et al. 2009 \\
$----$     & 1.82$\pm$0.06$^{*}$ & Kulkani et al. 2003 \\
SGR 1627     &1.87$\pm$0.18  & Esposito et al. 2009a, b \\
SGR 0501    & 6.3$\pm$1.7  & G\"{o}\v{g}\"{u}\c{s} et al. 2010 \\
PSR J1622      &$>$2.35$\pm$0.08 &  Levin et al. 2010\\
$----$     & $>$2.6$\pm$0.6$^{*}$ & Levin et al. 2010\\
1E 2259   &32$\pm$10 & Dib \& Kaspi 2014 \\
CXOU J1714   &2.1$\pm$0.9   & Sato et al. 2010 \\
$----$     & 2.2$\pm$0.9$^{*}$ & Halpern \& Gotthelf 2010b\\
$----$     &1.7$\pm$0.5$^{*}$ & Halpern \& Gotthelf 2010b\\
Swift J1834 &1.08$\pm$0.04   & Kargaltsev et al. 2012 \\
\hline\\
\end{tabular}\\
\label{tb:braking index}
\end{table}%
Since the first term in the right side of Eq.(10) is a constant, the
error of $n$ is solely determined by the variations of $\frac{\nu}{\dot{\nu}t_{\rm SNR}}$
(or of $\frac{P}{\dot{P}t_{\rm SNR}}$), the second term in the right side of Eq. (10), and
the standard error $\Delta n$ of $n$ can be expressed as the form of a standard propagated error\footnote{There is a relation between $\Delta{n}$ and $\Delta\left(\frac{\nu}{\dot{\nu}t_{\rm SNR}}\right)$,  $|\Delta{n}|=|\Delta\left(\frac{\nu}{\dot{\nu}t_{\rm SNR}}\right)|=[\left(\frac{\Delta \nu}{\dot{\nu} t_{\rm SNR}}\right)^{2}+\left(\frac{\nu \Delta t_{\rm SNR}}{\dot{\nu}t_{\rm SNR}^{2}}\right)^{2}+\left(\frac{\nu \Delta \dot{\nu}}{\dot{\nu^{2}}t_{\rm SNR}}\right)^{2}]^{1/2}$.}.
From Table 4, all the values of $n$ deviate from
1 and 3. Five magnetars have smaller
braking indices, $1<n< 3$, while the other three have larger braking
indices $n> 3$, which imply that neither pure PWR model nor pure MDR model
can account for the actual braking of magnetars.
In order to investigate the relation of $\tau_{\rm c}$
and $t_{\rm SNR}$,  we plot Log$_{10}(t_{\rm SNR}/\tau_{\rm c})$
versus Log$_{10}B$ for nine magnetars in Fig. 1.
\begin{figure}
\centerline
{
 \hspace{2mm}\epsfig{file=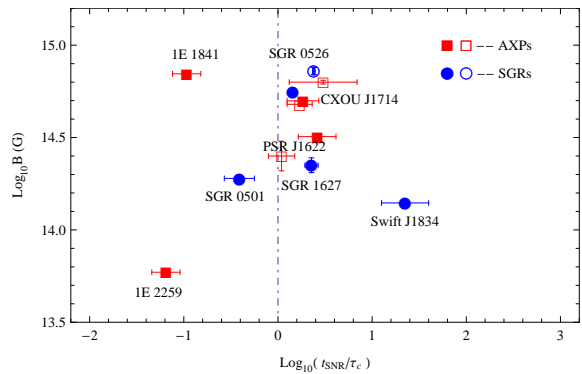,width=76mm,angle=-360}
}
\caption{Log$_{10}(t_{\rm SNR}/\tau_{\rm c})$ versus Log$_{10}B$.
The dot-dashed line corresponds to $t_{\rm SNR}= \tau_{\rm c}$.
Blue circles and red squares represent SGRs and AXPs,
respectively. Data for empty circles and squares are from Table 3.
For some sources (e.g., SGR 1627), the errors in data-points are smaller than the size of the symbols.}
 \label{evsb}
\end{figure}

In Fig. 1, magnetars with $n> 3$ are on the left of the dashed line whereas those with
$1<n<3$ are on the right. The large discrepancy between
$t_{\rm SNR}$ and $\tau_{\rm c}$ of a magnetar can be
naturally explained as follows: Inserting $P/\dot{P}=-\nu/\dot{\nu}=
2\tau_{\rm c}$ into Eq. (10) gives $\tau_{\rm c}/t_{\rm SNR}=(n-1)/2$;
if $n> 3$, then $\tau_{\rm c}> t_{\rm SNR}$, and a magnetar looks
``older'' than its true age (e.g., 1E 1841). In this
case, the larger the braking index is, the larger the disparity between
$\tau_{\rm c}$ and $t_{\rm SNR}$, as illustrated by
1E 2259. If $1<n<3$, then $\tau_{\rm c}<t_{\rm SNR}$,
and a magnetar appear ``younger'' than it is,  the smaller the braking
index is, the larger the age-disparity.
\subsection{Case of $n> 3$  \label{subsec:highindex}}
 The influence of the magnetic field evolution on the braking
indices of radio-pulsars has been well investigated in previous
studies (e.g., Geppert \& Rheinhardt 2002; Aguilara et al. 2008; Pons
\& Geppert 2007; Pons et al. 2009; Vigan\`{o}\& Pons 2012; Vigan\`{o}
et al. 2012). In this subsection, we focus on three high
braking indices ($n$=32$\pm$10, 13$\pm$4 and
6.3$\pm$1.7 for 1E 2259, 1E 1841 and SGR 0501, respectively). In order to
explain why the three magnetars possess large braking indices,
we consider the influences of  crust magnetic field decay on a
magnetar's spin-down, and refer to the works of Pons et al
(2009) and Vigan\`{o} et al. (2013).

According to Pons et al.(2012), the spin evolution of an isolated NS
is dominated by the magnetic field anchored to the star's
solid crust, and a varying crustal magnetic field
yields the braking index
\begin{equation}
n=3-4\frac{\dot{B}_{\rm d}}{B_{\rm d}}\frac{\nu}{\dot{\nu}}=3-4\frac{\dot{B}_{\rm p}}{B_{\rm p}}\frac{\nu}{\dot{\nu}}=3-4\frac{\tau_{c}}{\tau_{B}},
\label{11-equation}
\end{equation}
where $B_{\rm d}$ is the surface dipole magnetic field,
$B_{\rm p}=2B_{\rm d}$ is the surface magnetic field at the pole, and
$\tau_{B}=B_{\rm p}/\dot{B_{\rm p}}=B_{\rm d}/\dot{B_{\rm d}}$ is the
timescale on which the dipole component of the magnetic field evolves.
From Eq.(11), a decreasing surface dipole magnetic field, $B_{\rm d}$,
results in a decaying dipole braking torque, then the effective
braking index increases to a high value of $n> 3$.

In order to reconcile the measured high braking indices for three
magnetars with their (estimated) surface field decaying-rates, it is
necessary to introduce the Hall induction equation describing
the evolution of crust magnetic field in the following form:
\begin{equation}
\frac{\partial \vec{B}}{\partial t}=-\nabla\times\left[\frac{c^{2}}{4\pi\sigma}\nabla\times (e^{\nu}\vec{B})+\frac{c}{4\pi en_{e}}[\nabla\times(e^{\nu}\vec{B})]\times\vec{B}\right],
\label{12-equation}
\end{equation}
where $\sigma$ is the electric conductivity parallel
to the magnetic field, $e^{\nu}$ is the relativistic
redshift correction, $n_{e}$ is the number of electrons per unit volume,
and $e$ the electron charge.  On the right-hand side, the
first and second terms account for Ohmic dissipation
and Hall effect (which redistributes the energy of the
crustal magnetic field), respectively. As to the above differential
equation, Aguilara et al. (2008) presented an analytic solution
\begin{equation}
\frac{dB_{\rm p}}{dt}=-\frac{B_{\rm p}}{\tau_{\rm Ohm}}-
\frac{1}{B_{\rm p}^{i}}\frac{B_{\rm p}^{2}}{\tau_{\rm Hall}},
\label{13-equation}
\end{equation}
where $\tau_{\rm Ohm}$ is the Ohmic dissipation timescale, and
$\tau_{\rm Hall}$ is the Hall timescale corresponding to an
initial surface dipole magnetic field at the pole, $B_{\rm p}^{i}$. During a NS's early
magneto-thermal evolution stage, corresponding to the crust
temperature $T\geq 10^{8}$\,K,  the average
value of $\tau_{\rm Ohm}$ in the crust is about 1 Myrs (Pons
\& Gepport 2007). Since both the magnetic field strength and
the density vary over many orders of magnitude in NSs' crusts,
the average Hall timescale is roughly restricted as
$\tau_{\rm Hall}\sim \tau_{\rm Ohm}/(1-10)B_{13}$, where $B_{13}$ is the
dipole magnetic field at the pole in units of $10^{13}$\,Gauss (Pons
\& Gepport 2007).  We choose $\tau_{\rm Hall}\sim 5\times 10^{2}-5\times 10^{5}$\,yrs, similar as in
Vigan\`{o} et al. (2013), and $\tau_{\rm Ohm}\sim 10^{6}$\,yrs for the three magnetars.

To estimate the values of $B_{\rm d}$ for the three
magnetars, we refer to Vigan\`{o} et al. (2013):
\begin{enumerate}
\item  By introducing the state-of-the-art kinetic coefficients and
considering the important effect of the Hall term, Vigan\`{o} et al.
(2013) presented the updated results of 2D simulations of the
NS magneto-thermal evolution, and compared their results with a data
sample of 40 sources including magnetars 1E 2259,
1E 1841 and SGR 0501. It was found that by only changing the initial magnetic fields,
masses and envelope compositions of the NSs, the phenomenological diversity
of magnetars, high-B radio-pulsars, and isolated nearby NSs can be well
explained in their theoretical models.

\item These sources have best X-ray spectra by $Chandra$
or $XMM$-$Newton$, from which the luminosities and temperatures can be
obtained, their well-established associations, and known timing
properties with the characteristic age and the alternative estimate
for the age.  Similar as in this work, Vigan\`{o} et al. (2013)
selected their SNRs' ages $\tau_{k}$ as the actual age estimates for
the three magnetars  (for 1E 2259, 1E 1841 and SGR 0501,
$\log\tau_{k}$ (yr) $\simeq 4.0-4.3$, $2.7-3.0$, and 4,
respectively, corresponding to $\tau_{k}$ are 10$-$20, 0.5
$-$1.0, and 10 kyrs, respectively). For generality, the Hall
evolution timescales for the crustal fields have been arbitrarily
selected as $10^{3}$, $10^{4}$, $10^{5}$, and $5.0
\times10^{5}$\,yrs, respectively (Vigan\`{o} et al. 2013).

\item Comparing the cooling curves with iron envelopes for $B_{\rm p}^{i}=
10^{15}$ Gauss, 1E 1841 and SGR 0501 could be born with even higher magnetic
field of a few $10^{15}$\,Gauss, which can provide a strong hard X-ray
(20$-$200\,keV) emission with a total luminosity $\sim10^{36}$\,erg s$^{-1}$
(Vigan\`{o} et al. 2013).

\item Comparing the cooling curves for the group of NSs with $B_{\rm d}^{i}
\sim 1-5\times 10^{14}$\,Gauss suggests that 1E 2259 may be born with $B_{\rm p}^{i}\sim 3\times 10^{14}$\,Gauss,
needed to reconcile the observed timing properties and the persistent soft X-ray luminosity.

\end{enumerate}

According to the above arguments, we assume
$B_{p}^{i}=3.0\times 10^{15}$\,Gauss for 1E 1841 and SGR 0501,
and $B_{p}^{i}\sim 3\times 10^{14}$\,Gauss for 1E 2259.
Inserting the values of $B_{\rm p}^{i}$, $B_{\rm p}$, $\tau_{\rm Ohm}$, and $\tau_{\rm Hall}$
into Eq.(13), we can obtain the estimated values of the surface magnetic
field decay-rates $dB_{\rm p}/dt$. By solving Eq.(13), we obtain the relation of $dB_{\rm p}/dt$ and the field evolution
timescale $t$ for the three magnetars, as shown in Fig. 2.
\begin{figure}
\centerline
{
 \hspace{2mm}\epsfig{file=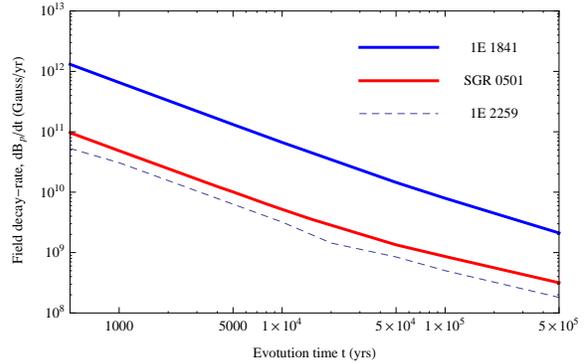,width=76mm,angle=-360}
}
\caption{The relation of $dB_{\rm p}/dt$ and $t$ for three magnetars with $n>3$.
The initial values of $B_{\rm p}$ are $1.4\times 10^{15}$, $3.8\times 10^{14}$, and
 $1.18\times 10^{14}$\,Gauss for 1E 1841, SGR 0501, and 1E 2259, respectively.}
 \label{evsb}
\end{figure}
From Fig.2, $dB_{\rm p}/dt$ decreases with increasing $t$. Due to the smallest value
of $dB_{\rm p}^{i}$, the values of $B_{\rm p}/dt$ for 1E 2259 are lower than those of
the other two magnetars.  Assuming $t=\tau_{k}$ (Vigan\`{o} et al. 2013), the current
values of $dB_{\rm p}/dt$ of three magnetars are calculated as follows: $dB_{\rm p}/dt=
-(1.32\times10^{12}-6.55\times10^{11}$\,Gauss\,yr$^{-1}$ for 1E 1841, $dB_{\rm p}/dt=
-(5.19\times10^{9}$\,Gauss\,yr$^{-1}$ for SGR 0501, and  $dB_{\rm p}/dt=
-(4.76\times10^{9}-2.44\times10^{9}$\,Gauss\,yr$^{-1}$ for 1E 2259. SGR 0501 has the largest
value of $dB_{\rm p}/dt$ because it has the smallest age $t=\tau_{k}=0.5-1.0$ kyrs.
Thus, the values of timescale $\tau_{B}=B_{\rm p}/\dot{B_{\rm p}}$ can also be estimated as follows:
for 1E 1841, $\tau_{B}=1.07-2.14$ kyrs, corresponding to $t=\tau_{k}=0.5-1.0$\,kyrs,
for SGR 0501, $\tau_{B}=73.2$ kyrs, corresponding to $t=\tau_{k}=10.0$\,kyrs, and for 1E 2259,
$\tau_{B}=24.8-48.4$ kyrs, corresponding to $t=\tau_{k}=10.0-20.0$\,kyrs.

Assuming the surface dipolar field decay is the only factor
 influencing magnetar braking indices, by combining the values of $dB_{\rm p}/dt$
with Eq.(11), we  plot the diagrams of $n$ versus $\log t$ for three
magnetars, shown as in Fig.3.
\begin{figure}
\centerline
{
 \hspace{2mm}\epsfig{file=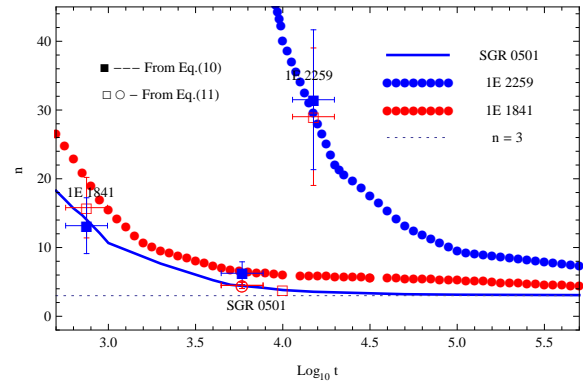,width=76mm,angle=-360}
}
\caption{The relation of $n$ and $t$ for three magnetars with $n>3$.
}
 \label{evsb}
\end{figure}
In Fig.3, the solid-squares denote the values of $n$ obtained from Eq.(10) , and
the empty-squares denote the values of $n$ obtained from Eq.(11) and the age-values presented in Vigan\`{o} et al.
(2013). The detailed calculations and comparisons are as follows:
\begin{enumerate}
\item For 1E 1841, when $t=\tau_{k}=0.5-1.0$\,kyrs, which is consistent with the
SNR age estimated by this work, its braking index $n=11.6-20.2$, which is
comparable to the measured value of $n\sim$9$-$17 (see Table 4) in the magnitude.

\item  For 1E 2259, when $t=\tau_{k}=10.0-20.0$\,kyrs, which is also consistent with the SNR age estimated by this work,
its braking index $n$=20$-$40, which approaches to the measured value of $n\sim$ 22$-$42 (see Table 4) in the magnitude.

\item
For SGR 0501, when $t=\tau_{k}=10.0$\,kyrs, its braking index $n=$3.8, which is apparently
smaller than the measured value of $n\sim$ 4.6$-$8.0 (see Table 4). This discrepancy may be caused by: 1)
the in1tial dipolar magnetic field estimated is too low;  2) the age estimate is too large (compared with
$t=t_{SNR}=4.0-7.7$\,kyrs in this work), and 3)the decay of the magnetospheric braking torque, e.g.,
the surface dipolar field decay cannot be the only factor
 influencing magnetar braking indices. Here, we insert $t=t_{SNR}=4.0-7.7$\,kyrs into Eq.(11), then get $n\sim$4.1$-$5.0 (as denoted in the empty-circle in Fig.3), which is slightly larger than the value of $n=$3.8 by Vigan\`{o} et al.(2013).
\end{enumerate}

Although we have attributed the high-value braking indices of the three magnetars to the decay
of the crustal surface magnetic field, there is the possibility of the decay of the magnetospheric braking torque,
which can also result in the braking indices  $n> 3$. Below we will discuss
the spin-down evolution for the three sources.

SGR 0501 was firstly discovered by the Swift $\gamma$-ray
observatory (Barthelmy et al. 2008). From the \textit{RXTE/PCA, Swift/XRT,
Chandra/ACIS-S}, and \textit{XMM-Newton/PN} observations spanning a short range of MJD
54700$-$54950, its magnitude of $\dot{\nu}$ decreased from
$-2.03(8)\times10^{-13}$ to $-1.75(6)\times10^{-13}$\,Hz\,s$^{-1}$
\citep{gwk08, gog10}. If this decrease in
$\dot{\nu}$ is solely caused by the decay of $B_{\rm p}$, the mean surface dipolar
field decaying-rate of $dB_{\rm p}/dt$ is
$-(4.6\pm0.8)\times10^{13}$\,Gauss\,yr$^{-1}$ (for a rotator in
vacuum, $B_{\rm p}\simeq 6.4\times10^{32}(-\dot{\nu}/\nu^{3})^{1/2}$\,Gauss), corresponding to
the braking index $n\sim 63.1-88.3$ obtained from Eq.(11).

Recently, Dib \& Kaspi (2014) presented a summary of
the long-term evolution of various properties of five AXPs
including 1E 1841 and 1E 2259, regularly monitored
with \textit{RXTE} from 1996 to 2012. Influenced by frequent
glitches and magnetar outbursts, 1E 1841 and 1E 2259
exhibited large fluctuations in their spin-down parameters $\nu$ and
$\dot{\nu}$. However, both AXPs obviously exhibited net decrease in $\dot{\nu}$.
For 1E 2259, its $\dot{\nu}$ decreased from
$-1.020(17)\times10^{-14}$ to $-0.96748\times10^{-14}$\,Hz\,s$^{-1}$ during an
interval of MJD 50355$-$54852\citep{dk14}, which corresponds to
$n\sim 1404.2-2653.8$ according to Eq.(11).
For 1E 1841, its $\dot{\nu}$ decreased from
$-2.9954(7)\times10^{-14}$ to $-2.944(11)\times10^{-13}$\,Hz\,s$^{-1}$ during a
interval of MJD 51225$-$55903\citep{dk14}, which corresponds to
$n\sim 9.8-15.6$.

From the analysis above, apart from 1E 1841, the estimated values of $n$ from the surface
dipolar magnetic field decay are far larger than the values of the measured mean braking indices
for magnetars 1E 2259 and SGR 0501. We suppose that these large and rapid decreases in $\dot{\nu}$ were explained as
the effects of timing-noise torque and outbursts \citep{gog10}, instead
of as a secular dipole magnetic field decay. An actual decay of magnetospheric torque may explain  the high braking index for a magnetar,
but could result in an over-estimate of the surface dipolar field decay rate.

Many authors (e.g., Harding et al 1999; Spitkovsky
2006; Beskin \& Nokhrina 2007; Tasng \& Gourgouliatos 2013; Antonopoulou
et al. 2015) investigated the external magnetospheric mechanisms that can
explain larger decrease in $|\dot{\nu}|$ during the pulsar spin-down.
The rotation of a pulsar induces an electric field accelerating
charges off the star's surface. The magnetosphere is expected to be
filled with cascade plasma, which will screen the electric field
along the magnetic field lines (Goldreich \& Julian 1969).  The
magnetospheric current, $\vec{J}$, flowing in the magnetosphere
and closing under the polar-cap surface, provides an additional braking torque,
\begin{equation}
-\vec{T}_{\rm j}=\frac{1}{c}\int[\vec{r}, [\vec{J}_{s}, \vec{B}_{\rm p}]]dS~.
\label{14-equation}
\end{equation}
This braking torque on the star from the magnetospheric current
is comparable to $-T_{\rm MDR}$, the braking torque on the star for MDR model
\footnote{In the simplest approximation for pulsar spin-down in MDR model,
the braking torque on the star is $-T_{\rm MDR}=\frac{B^{2}_{p}R^{6}
\Omega^{3}}{6c^{3}}{\rm sin}^{2}\chi$, here $R$ is the stellar radius, and $\chi$
is the inclination angle, i.e., the angle between rotational and magnetic axes.} From Eq. (14), a decreasing
magnetospheric current $\vec{J}$ produces a decaying $-\vec{T}_{\rm j}$, which
will cause a decrease in $\dot{\nu}$.  For an net decay of toque due to
the decrease in $\vec{J}$, the relative decrease in the spin-down rate is estimated as
$\Delta T_{\rm j}/T_{\rm j}\simeq \Delta\dot{\Omega}/\dot{\Omega}=
\Delta\dot{\nu}/\dot{\nu}$ assuming the momentum of inertial $I$, and the angle
between rotational and magnetic axes, $\chi$ are constants.
Note that, a precise calculation of $\dot{\Omega}$ (or $\dot{\nu}$) depends on solving the
Euler dynamics equations that are very complicate and uncertain (Beskin
\& Nokhrina 2007).

It is also possible that a decrease in $\chi$ can cause a
decrease in spin-down rate, as observed. During the long-term
observations, these three magnetars with SNRs experienced frequent outbursts and glitches.
\citep{gwk08, gog10, dk14}.  During post-outburst or glitch recoveries, the
stellar platelets move towards the rotational axes.  The
magnetic field lines anchored to the crusts follow the motion,and the
inclination angle $\chi$ decreases (Antonopoulou et al. 2015).
A decrease in $\chi$ will result in a net decay of the magnetospheric
braking toque. Thus, the overall effect can be a decrease in
$\dot{\nu}$. For a small decrease of $\chi$, the relative decrease in
$\dot{\nu}$ is estimated as $\Delta
T_{\rm dip}/T_{\rm MDR}\simeq \Delta \chi/{\rm tan}\chi\simeq
\Delta\dot{\nu}/\dot{\nu}$ in MDR model, or as $\Delta T_{\rm ff}/T_{\rm ff}
\simeq \Delta \chi{\rm \sin}2\chi/(1+{\rm \sin}^{2}\chi)\simeq
\Delta\dot{\nu}/\dot{\nu}$, here $T_{\rm ff}$ is the braking torque
for the force-free magnetosphere\footnote{If the magnetosphere
is filled with abundant plasma  it can be considered in the
force-free regime, where the induced electric field along field
lines disappears everywhere except in the acceleration zones above
polar caps and regions where plasma flow is required, the braking
torque on the star, $-T_{\rm ff}= \frac{B^{2}_{p}R^{6}\Omega^{3}}
{c^{3}}(1+{\rm \sin}^{2}\chi)$ (Spitkovsky 2006).}(Spitkovsky 2006;
Antonopoulou et al. 2015). If the external magnetospheric mechanisms
(whether due to the change in the current or due to the change in the inclination angle)
dominate the spin-down evolutions for three magnetars, the order of magnitude of
the relative decreases in the braking torque is about $|\Delta T/T|\approx
|\Delta \dot{\nu}/\dot{\nu}|\sim 10^{-1}-10^{-2}$, as observed in 1E 2259, 1E 1841 and SGR 0501.

In summary, we have quantitatively analyzed several physical
processes (the decay of surface dipolar field, the decline of
the current, and the decrease in the inclination angle), possibly
causing $n>3$ for magnetars. A common characteristic of these
processes is that they can result in a decay of the braking torque
on the star. Here we ignore the effects of the stellar multipole magnetic moments on
the magnetar braking indices, because the higher-order magnetic moments decay rapidly
with the distance to stellar center.
\subsection{Case of $1<n< 3$ \label{subsec:windluminosity}}
In previous works, many authors proposed various models to
explain why the observed braking indices of pulsars $1<n<3$, e.g.,
neutrino and photon radiation \citep{phh82}; the combination of dipole
radiation and the propeller torque applied by debris-disk \citep{ab06, cl06};
frequent glitches, as well as magnetosphere currents (Beskin \& Nokhrina 2007). In this work, the
lower braking indices of $1<n<3$ for five magnetars with SNRs are
attributed to the combination of MDR and wind-aided braking.  Recently, the
extended emission observed from 1E 1547 and Swift J1834
are proposed to be magnetism-powered pulsar-wind-nebulas (PWNs), which may
accelerate plasma to very high energy and radiate high-energy photons (see
Kargaltsev et al. 2012, and references therein).  The possible existence of
magnetar PWNs lends a potential and direct support to our assumption of
wind-aided braking.

With the aid of strong magnetar wind, the closed magnetic-field lines
are combed out at the extended Alfv\'{e}n  radius $r_{\rm A}$, where
the magnetic energy density equals the particle kinetic-energy density.
The braking toque resulting from tangential acceleration of outflowing
relativistic plasma is $-T_{\rm wind}\approx \dot{M}\Omega r^{2}_{\rm A}
$, where $\dot{M}$ is the total mass-loss rate\citep{mdn85}. The
accelerated outflowing plasma will carry away the star's angular momentum,
and thus the spin-down rate of the magnetar will increase.
The added braking toque due to a magnetar wind extends the effective
magnetic braking ``lever arm '', 
causing the magnetar braking index to decrease, i.e., $n<3$.

In the scenario of wind-aided braking, the value of the effective braking
index $n$ of a magnetar depends on the competition of the magneto-dipole
braking toque $-T_{\rm dip}$ and wind-braking toque $-T_{\rm wind}$. If
$|T_{\rm wind}|\gg |T_{\rm dip}|$, the value of $n$ will be close
to 1, and if $|T_{\rm dip}|\gg |T_{\rm wind}|$, the value of $n$ will be close to 3.
Unlike the persistent soft X-ray luminosity of a magnetar $L_{\rm X}$,
the persistent magnetar wind luminosity $L_{\rm W}$ cannot be obtained from observations.
Moreover, it's very difficult to determine $L_{\rm W}$ of a magnetar theoretically,
due to the lack of a detailed magnetar-wind mechanism.
The well-known magnetar wind-braking model proposed by Harding et al. (1999)
implies a braking index of
\begin{equation}
n=~3-~\frac{2\nu}{\mid\dot{\nu}\mid}\frac{L_{\rm W}^{1/2}B_{\rm d}R^3}{2I\sqrt{6c^{3}}}~,
\label{15-equation}
\end{equation}
where we employ standard values of $R=10 \rm km$ and moment of inertia
$I=10^{45}$\,g\,cm$^{2}$. Note that, by definition, the inferred value of
$n$ is always within the range of 1$-$3.  In the above expression, $n$ is an
effective braking index, thus $L_{\rm W}$ is the effective wind-luminosity,
and $B_{\rm d}$ is the effective surface dipolar field. However, due to
lack of the detailed information of the effective $n$, $B_{\rm d}$ and
$L_{\rm W}$, the importance of Eq.(15) has not been recognized in previous
works.  From the above expression, the magnetar wind luminosity can be estimated as
\begin{equation}
L_{\rm W}= \frac{6(3-n)^{2}\dot{\nu}^{2}I^{2} c^{3}}{\nu^{2}B_{\rm d}^{2}R^{6}}.
\label{16-equation}
\end{equation}
By using Eq.(16), we estimated possible wind luminosity of magnetars,
$L_{\rm W}$, as listed in Table 5. Here we assume that the effective value of
surface dipolar field to be the observed value of $B_{\rm d}$ for convenience.
\begin{table*}
\begin{center}
\centering
\begin{minipage}{178mm}
\caption{The relations among $L_{\rm rot}$, $L_{\rm X}$ and $L_{\rm W}$ of magnetars with different
braking indices. Five sources in the upper part of Table 5 have braking indices $1<n<3$, while
three sources in the lower part have braking indices $n>3$. All the date $L_{\rm X}$(2-10\,keV) in
quiescence are cited from McGill SGR/AXP Online Catalogue. }
\label{tb:parameters-6}
\begin{tabular}{cccccccc}\\
\hline 
Name &$L_{\rm rot}$    &$L_{\rm X} $   &$L_{\rm W}$     &$L_{\rm W}/L_{\rm rot}$&$L_{\rm W}/L_{\rm X}$&$L_{\rm X}/L_{\rm rot}$  \\
         &($\rm erg~s^{-1}$) &($\rm erg~s^{-1}$) &($\rm erg~s^{-1}$)&               &             &  \\
 \hline  \\
SGR 1627$^{*}$  & $4.3(9)\times10^{34}$ & $3.6\times10^{33}$ & (2.304$\pm$0.997)$\times10^{35}$ &5.36$\pm$2.58 &64.0$\pm$27.7 & 0.08$\pm$0.02\\
Swift J1834$^{*}$  & $2.1\times10^{34}$ &$<$ 0.84$\times10^{31}$ & (3.16$\pm$0.17)$\times10^{35}$& 15.03$\pm$0.78
&37583.3$\pm$1964.3 &$<$0.0004\\
CXOU J1714 &$4.51\times10^{34}$&$5.6\times10^{34}$&(1.53$\pm$1.52)$\times10^{35}$& 3.39$\pm$3.37 &2.73$\pm$2.71 & 1.242\\
 $---^{\dag}$&4.15$\times10^{34}$& 5.6$\times10^{34}$&(1.15$\pm$1.57)$\times10^{35}$ &2.77$\pm$3.78& 2.05$\pm$2.80& 1.35\\
 $---^{\dag}$&7.41(35)$\times10^{34}$& 5.6$\times10^{34}$&(5.36$\pm$4.26)$\times10^{35}$ &7.23$\pm$5.76& 9.57$\pm$7.61& 0.76$\pm$0.04\\
SGR 0526 &2.87(7)$\times10^{33}$&$1.89\times10^{35}$&(4.13$\pm$0.55)
$\times10^{33}$ &1.44$\pm$0.19& 0.002$\pm$0.003& 65.85$\pm$1.61 \\
$---^{\dag}$&4.9(4)$\times10^{33}$& 1.89$\times10^{35}$&(2.75$\pm$0.53)$\times10^{34}$ &5.61$\pm$1.18& 0.146$\pm$0.028& 38.57$\pm$3.15\\
PSR J1622 &8.3(5)$\times10^{33}$&$4.4\times10^{32}$&(1.47$\pm$0.40)$\times10^{34}$
&$1.77\pm$0.49 &33.41$\pm$9.09& 0.053$\pm$0.003\\
$---^{\dag}$&(7.0$\pm$2.4)$\times10^{33}$& 4.4$\times10^{32}$&(4.87$\pm$4.13)$\times10^{33}$ &0.69$\pm$0.58& 11.1$\pm$9.4& 0.063$\pm$0.022\\
\hline\\
1E 2259$+$586 &  $ 5.6\times 10^{31}$  & $ 1.7\times10^{34}$    & .....& .....& .....& 303.57\\
1E 1841$-$045 &  9.9$\times 10^{32}$ & $ 1.84\times 10^{35}$ & .....& .....& .....& 185.85\\
SGR 0501$+$4516   & $ 1.2\times10^{33}$  & $8.1\times 10^{32}$ & .....& .....& .....& 0.67\\
\hline\\
\end{tabular}\\
\noindent{ \\}
\noindent{Footnote:$^{*}$ According to the ``fundamental plane''(Rea et al. 2012) for radio magnetars with $L_{\rm rot}> L_{\rm X}$, this source should have radio emission, however no radio emission was reported to date.}
\end{minipage}
\end{center}
\end{table*}

The rotation-energy loss rate is defined as $L_{\rm rot}=I\Omega\dot{\Omega}=4\pi^{2}I\nu\dot{\nu}=
-4\pi^{2}I\dot{P}P^{-3}$, here $I=10^{45}$\,g\,cm$^{2}$ for all sources.
From Table 5, the value of $L_{\rm X}$ is obviously unrelated to that of
$L_{\rm rot}$, this is because magnetar X-ray emission is powered by
magnetic field free energy.  With the exception of SGR 0526, magnetars'
wind luminosities are bigger than their soft X-ray luminosities.
In addition, 1E 2259 possesses a relatively large ratio
of $L_{\rm X}/L_{\rm rot}\sim 303.57$ (see in Table 5), which is higher
than those of all magnetars. The large ratio of $L_{\rm X}/L_{\rm rot}$ could be served as a
strong hint of surface dipolar field decay for this source, as discussed in Sec.3.2.

Recently, Tong et al. (2013) proposed a multipole field-powered
wind model for magnetars. The authors assumed that all the high luminosity
magnetars with $L_{\rm X}> L_{\rm rot}$ spin down via wind braking, and
transient magnetars  with $L_{\rm X}<L_{\rm rot}$ spin down via magnetic dipole
braking. Based on a simple assumption of $L_{\rm W}=L_{\rm P}=10^{35} \,
\rm erg\,s^{-1}$ ($L_{\rm P}$ is the escaping particle luminosity).
They concluded that, for typical AXPs and SGRs, the dipole magnetic field
in the case of wind braking is about ten-times lower than that of magnetic-
dipole braking, and a strong dipole magnetic field may be no longer necessary for
some sources and so on. However, when estimating $B_{\rm d}$, one should be cautious taking  into account effective braking indices.
\section{Summary and Discussion \label{sec:sad}}
Considering the possible SNR associations and their ages and assuming a constant
braking index over the magnetar's lives, we have estimated the average values of braking
indices,for the eight magnetar, and discussed the possible origins of the braking indices. Our main conclusions are as follows.
\begin{enumerate}
\item  Based on possible timing solutions of magnetars and
updated measurements of their associated SNRs, we estimate the
braking indices $n$ of eight magnetars, and find that they cluster in a range of $n\sim 1-42$.

\item  We attribute the higher braking indices, $n>3$, of
1E 2259, 1E 1841 and SGR 0501 to the surface dipolar
field decay, or to the decay of the external magnetospheric braking torque.
In the former case, the estimated
dipolar field decay rates, based on the updated magneto-thermal
evolution models, are compatible with the measured values of $n$ for
these three sources.

\item  We attribute the lower mean braking indices, $1<n< 3$, of other five
magnetars (SGR 0526, SGR 1627, PSR J1622, CXOU J1714 and
Swift J1734) to a combination of MDR and wind-aided
braking, and estimate their possible wind luminosities $L_{\rm W}$.

\end{enumerate}

At last, we remind that, in order to
derive an analytical expression for estimating the magnetar's true age,
the braking index is assumed to be a constant.
However, the inferred quantities (e.g., $B_{\rm d}$ and $\tau_{\rm c}$) of a magnetar
strongly depend on the current values of $P$ and $\dot{P}$ (or of $\nu$
and $\dot{\nu}$) that\textbf{ can} vary with time.  There always exist
observational errors when estimating $t_{\rm SNR}$ for magnetars. Thus, for
some sources, their constrained values of $n$ are subject to substantial uncertainties, and
will be modified with future observations. We expect that more
observations of magnetars and their SNRs will help in further constraining the
magnetar braking mechanisms.

\section*{Acknowledgments}
We thank anonymous referees for carefully reading the manuscript and
valuable comments that helped improve this paper substantially. We
also thank Prof. Yang Chen for discussions, Dr. Rai Yuen
for smoothing out the language, and  Dr. Cristobal Espinoza for useful
suggestions on braking indices of several radio pulsars. This work was supported by Xinjiang
Natural Science Foundation No.2013211A053, by Chinese National
Science Foundation through grants No.11173041,11173042, 11273051,
11373006 and 11133001, National Basic Research Program of China
grants 973 Programs 2012CB821801, the Strategic Priority Research
Program ``The Emergence of Cosmological Structures'' of Chinese
Academy of Sciences through No.XDB09000000, and the West Light
Foundation of Chinese Academy of Sciences No.2172201302.

\label{lastpage}

\end{document}